\documentclass{sig-alternate}
\usepackage[ruled,vlined]{algorithm2e}
\usepackage{graphicx}
\usepackage{url}
\usepackage{paralist}
\usepackage{tabularx}
\usepackage{balance}
\usepackage{multirow}
\usepackage{multicol}
\usepackage{pbox}
\usepackage{mathtools}
\usepackage[TABBOTCAP]{subfigure}
\usepackage{algorithmic}
\usepackage{graphicx}
\usepackage{paralist}
\usepackage{tabularx}
\usepackage{url}
\usepackage{balance}
\usepackage{multirow}
\usepackage{multicol}
\usepackage{amsmath}
\usepackage{setspace}
\usepackage[bookmarks=false]{hyperref}
\usepackage{verbatim}
\usepackage{float}
\usepackage[normalem]{ulem}
\usepackage{xspace}
\usepackage[usenames]{color}
\usepackage{colortbl}
\usepackage{amssymb}
\usepackage{ifthen}
\usepackage{pifont}
\usepackage{tikz}

\include{macro}

\newcommand{\Fusionsp}{{\sc Fusion~}}
\newcommand{\Fusion}{{\sc Fusion}}

\newcommand\blfootnote[1]{%
  \begingroup
  \renewcommand\thefootnote{}\footnote{#1}%
  \addtocounter{footnote}{-1}%
  \endgroup
}

\begin{document}
\CopyrightYear{2016} 
\setcopyright{acmlicensed}
\conferenceinfo{ICSE '16 Companion,}{May 14 - 22, 2016, Austin, TX, USA}
\isbn{978-1-4503-4205-6/16/05}\acmPrice{\$15.00}
\doi{http://dx.doi.org/10.1145/2889160.2889269}

%

\title{Fixing Bug Reporting for Mobile and GUI-Based Applications}
%
%
%
%
%
\vspace{-0.3cm}
\numberofauthors{1} 
%
\author{
%
%
\alignauthor
Kevin Moran\\
\affaddr{Advisor: Denys Poshyvanyk}\\
\affaddr{Department of Computer Science, College of William \& Mary}\\
\affaddr{http://www.kpmoran.com}\\
\email{\{kpmoran, denys\}@cs.wm.edu}
}
\maketitle
\begin{abstract}
  Smartphones and tablets have established themselves as mainstays in the modern computing landscape.  It is conceivable that in the near future such devices may supplant laptops and desktops, becoming many users primary means of carrying out typical computer assisted tasks.  In turn, this means that mobile applications will continue on a trajectory to becoming more complex, and the primary focus of millions of developers worldwide.  In order to properly create and maintain these ``apps" developers will need support, especially with regard to the prompt confirmation and resolution of bug reports.  Unfortunately, current issue tracking systems typically only implement collection of coarse grained natural language descriptions, and lack features to facilitate reporters including important information in their reports.  This illustrates the \textit{lexical information gap} that exists in current bug reporting systems for mobile and GUI-based apps.  This paper outlines promising preliminary work towards addressing this problem and proposes a comprehensive research program which aims to implement new bug reporting mechanisms and examine the impact that they might have on related software maintenance tasks.
\end{abstract}
	
\vspace{-0.3cm}

\ccsdesc[500]{Software and its engineering~Software maintenance tools}
\ccsdesc[500]{Software and its engineering~Software testing and debugging}
\ccsdesc[500]{Software and its engineering~Maintaining software}
\ccsdesc[300]{Software and its engineering~Software evolution}
\ccsdesc[100]{Software and its engineering~Software configuration management and version control systems}
\printccsdesc
\vspace{-0.3cm}
\section{Introduction \& Motivation}
\label{sec:intro}
Mobile app development has become incredibly popular with more than 2 million mobile developers estimated to be actively creating and maintaining apps \cite{24MobilityReport}. However, the intense competition present in mobile application marketplaces like Google Play and the Apple App Store, means that if an app is not performing as expected,  due to bugs or lack of desired features, nearly half of users are less likely to use the app again and will abandon it for another with similar functionality \cite{app-abandonment}. This competition highlights the need 
\\
\\
\\
for developers to maintain high-quality, defect-free apps if they hope to capture user interest and market share.  Software maintenance activities are known to be generally expensive and challenging \cite{25Tassey:NIST} and one of the most important maintenance tasks is bug report resolution.  However, current bug tracking systems such as Bugzilla, Mantis, the Google Code Issue Tracker, the GitHub Issue Tracker, and commercial solutions such as JIRA rely mostly on unstructured natural language bug descriptions, with additional effort required to enhance reports with more useful information. The quality of the description mostly depends on the reporter's experience and attitude towards providing enough information. Therefore, the reporting process can be cumbersome, and the additional effort means that many users are unlikely to enhance their reports with extra information \cite{11Bettenburg:MSR08,31Davies:ESEM2014,32Bettenburg:ICSM08, 34Aranda:ICSE09}.   Additionally, developers face multiple other mobile-specific challenges in the issue resolution process such as frequent platform updates and API instability \cite{Linares-Vasquez:FSE2013}, platform and device fragmentation\cite{Han:WCRE2012}, and pressure for rapid releases\cite{market-pressure}.\blfootnote{This work is supported in part by the NSF CCF-1218129 and CCF-1525902 grants.}	

	 It is clear that the underlying \textit{task} that bug reporting systems must accomplish is to \textit{bridge the lexical knowledge gap between typical reporters of a bug and the developers that must resolve the bugs.} Bug and error reporting has been an active area of research in the software engineering community.  However, little work has been conducted to address the information gap, improve the lack of structure in the reporting mechanism, ease reporter difficulty for entering reproduction steps, and add corresponding support in issue tracking systems. Previous studies on bug report quality serve as further evidence of this problem.  In particular, these studies have indicated that (i) lack of information in reports contributes to non-reproducible reports\cite{4Joorabchi:MSR14}, (ii) information that developers find most useful (e.g. reproduction steps, stack traces, and test cases) are the most difficult items for reporters to provide \cite{3Bettenburg:FSE08}, and (iii) that information needs are highest at a bug's inception \cite{15Breu:CSCW10}.  
	 
	 These issues make it clear that current bug reporting mechanisms are not sufficient in supporting developers reproducing and fixing reported faults in software, especially for mobile GUI-based apps. However, the implications of the quality of these reports extend much further than this. Specifically, the current state and quality of bug reports cannot effectively be used to support costly software maintenance tasks.  While reproduction and fixing of bugs are important maintenance tasks, they are often the last steps in a complex maintenance cycle, with many connected tasks directly dependent upon the quality of incoming reports.  Two examples of difficult, and often costly tasks along this cycle are (i) the detection of duplicate bug reports to reduce the information burden on developers, and (ii) developer triaging which ensures that bug reports are assigned to developers who have the best expertise to quickly fix them.   Research to support these tasks faces many challenges, chief among them being the limited information typically contained within bug reports.  The rapidly evolving nature of software projects (especially mobile apps) represents another confounding factor for these techniques as bug reports and associated test cases often become outdated as software evolves.  \textit{By improving the bug reporting mechanism to include more detailed information, these maintenance tasks can be better supported and the reports and associated test cases could be made to evolve with a system.}  Considering all the current problems and challenges of current bug reports, it is clear that \textit{a paradigm shift in the how bugs are reported and subsequently leveraged to support software maintenance tasks is not only necessary, but a logical step forward in research}.  In the remainder of this paper, we introduce preliminary and proposed work towards improving bug reporting mechanisms for mobile and GUI-based apps, and outline new applications for using the created reports to support the maintenance tasks of developer triaging and duplicate report detection.
	\vspace{-0.25cm}
\subsection{Differentiation from In-Field Failure Reproduction}
	A body of work known as in-field failure reproduction \cite{26Jin:ISSTA13,18Jin:ICSE12} shares similar goals with our proposed approach. These techniques collect run-time information (e.g., execution traces) from instrumented programs that provide developers with a better understanding of the causes of an in-field failure, which will subsequently help expedite the fixing of those failures.  However, there are several key differences that illustrate how our research program improves upon the state of research and practice in this area.   
	\textit{First}, techniques regarding in-field failure reproduction rely on potentially expensive program instrumentation, which typically requires developers to modify code and introduce overhead.  Our proposed bug reporting mechanisms are completely automatic; our static and dynamic analysis techniques only need to be applied once for the version of the program that is released for testing.  Furthermore, the analysis process can be done without the need for instrumentation of programs in the field.  \textit{Second}, current in-field failure reproduction techniques require an oracle to signify when a failure has occurred (e.g., a crash) and this oracle must function in the field.  Our proposed research program approach is aimed primarily at supporting testers during the bug reporting process and is \textit{additionally} aimed at automatically generating natural language reports for well-established oracles, such as crashes, in-house. \textit{Third}, these techniques have not been applied to mobile apps and would most likely need to be optimized further to be applicable for the corresponding resource-constrained environment.

\vspace{-0.40cm}
\section{Preliminary Research}
\label{sec:preliminary-work}
	The design of the proposed research is inspired by promising preliminary results of recent work conducted in improving off-device bug reporting for Android apps, instantiated in an approach called \Fusionsp \cite{Moran:FSE2015}.  The tool is currently available for use online at \cite{appendix}.  This work illustrates the key idea behind the reporting mechanisms in the proposed research, namely that \textit{program analysis (and, as we propose later, statistical modeling) can be used to narrow the lexical information gap between reporters and developers}.  This idea will change the way that developers, testers, and end-users currently perceive bug reporting, as it allows testers to easily convey and describe problems, users to more easily engage developers in reporting problems and feature requests in applications, and developers to receive more detailed information and feedback than with any current state of the art issue tracking system.  The power of this approach can be further seen in the manner that high-level, adaptable, representations of execution scenarios are used to both guide users through the reporting process, and augment the bug reports with execution information that can be automatically replayed on different devices and platform versions.  To clearly illustrate our underlying key idea for the future of bug reporting in mobile and GUI-based apps, and highlight a subset of the opportunities for future work, we outline this preliminary work and briefly summarize promising evaluation results.  For more details, refer to \cite{Moran:FSE2015}.
	
	\Fusion's implements an \textit{Analyze} $\rightarrow$ \textit{Generate} workflow that corresponds to two major phases. In the \textit{Analysis Phase} \Fusionsp collects information related to the GUI components and event flow of an app through a combination of static and dynamic analysis.  Then in the \textit{Report Generation Phase}, \Fusionsp takes advantage of the GUI-centric nature of mobile apps to both auto-complete the steps to reproduce the bug and augment each step with contextual app information.   Videos of our prototype in use, complete with commentary, are available online \cite{appendix}.  
\vspace{-0.2cm}
\subsection{FUSION Analysis Phase}

    The \textit{Analysis Phase} collects all of the information required for the \textit{Report Generation Phase} operation. This first phase has two major components: 1) static analysis \emph{(Primer)}, and 2) dynamic program analysis \emph{(Engine)} of a target app. 
    The \textit{Analysis Phase} must be performed before each version of an app is released for testing or before it is published to end users.  Both components of the \emph{Analysis Phase} store their extracted data in the \Fusionsp database.

    \textbf{Static Analysis (Primer):} The goal of the \emph{Primer} is to extract all of the GUI components and their associated information from the app source code. Thus, this provides a universe of possible components within the domain of the app, and establishes traceability links connecting GUI components that reporters operate upon to code specific information such as the class or activity they are located within.  However, this preliminary approach utilizes only lightweight static analysis techniques and is not able to map every extracted GUI component to its handler in an activity. In future work we hope to leverage new advances in state-of-the-art static analysis techniques for Android apps \cite{Huang:ISSTA2015,Yang:ICSE2015}.


\textbf{Dynamic Analysis (Engine):} The \emph{Engine} is used to glean dynamic contextual information and enhance the data- base with both run-time GUI and app event-flow information.To do this it explores an app in a systematic manner, ripping and extracting run-time information related to the GUI components during execution including text, location, and cropped screenshots associated with the component.
 
    The \emph{Engine} performs this systematic exploration using modified subroutines from the {\tt UIAutomator} framework included in the Android SDK.  This systematic execution of the app is similar to existing approaches in GUI ripping \cite{Amalfitano:ASE2012,Azim:OOPSLA2013,Machiry:FSE2013}. However, using the {\tt UIAutomator} framework allows for detecting cases that are not captured in previous tools such as pop-up menus that exist within menus, internal windows, and the onscreen keyboard.  To explore the application we implemented our own version of a systematic depth-first search (DFS) algorithm for application traversal that performs click (tap) events on all the clickable components in the GUI hierarchy reachable using the DFS heuristic.  This type of exploration is intrinsically limited, potentially missing several important states in the execution of an application.  However, in future work we hope to explore promising recent approaches in GUI-based application testing\cite{Nguyen:TOSE2014}, as well as different heuristics for systematic exploration.
\vspace{-0.2cm}
\subsection{FUSION Report Generation Phase}  

                During the \emph{Report Generation Phase}, \Fusionsp aids the reporter in constructing the steps needed to reproduce a bug by making suggestions based upon the ``potential" GUI state reached by the declared steps.  This means for each step $s$, \Fusionsp infers the GUI state $GUI_s$ in which the target app should be by taking into account the history of steps.   For each step, \Fusionsp verifies that the suggestion made to the reporter is correct by presenting the reporter with contextually relevant screen-shots, where the reporter selects the screen-shot corresponding to the current action they want to describe.  The Fusion interface also collects general information about a bug such as a title, the device being used, and a brief high-level description.


\textbf{Auto-Completing Bug Reproduction Steps:} To facilitate the reporter in entering reproduction steps, we model each step in the reproduction process as an {\tt \{action, component\}} tuple corresponding to the action the reporter wants to describe at each step, (e.g., tap, long-touch, swipe, type) and the component in the app GUI with which they interacted (e.g.,``Name" textview, ``OK" button, ``Days" spinner).  Since reporters are generally aware of the actions and GUI elements they interact with, it follows that this is an intuitive manner for them to construct reproduction steps.  \Fusionsp allocates auto-completion suggestions to drop down lists based on a decision tree taking into account a reporter's position in the app execution beginning from a cold-start of the app. The first drop down list corresponds to the possible actions a user can perform at a given point in app execution.  When the reporter selects the \textit{type} option, we also present them with a text box to collect entered data.

	The second dropdown list corresponds to the component associated with the action in the step and presents information that helps reporters identify individual components by presenting information such as component-specific screenshots, and the relative location of the component on the screen.   To complete the step entry, the reporter selects a full-size contextual screen-shot corresponding to both the app state and the GUI component acted upon. After the reporter makes selections from the drop-down lists, they have an opportunity to enter additional information for each step (e.g., a button had an unexpected behavior) in a natural language text entry field.  The combo box interface can become cumbersome for reporters to use, particularly if there are many extracted components for a current app's screen, however we hope to explore novel UI designs in the future.

\textbf{Report Structure:} The Report presents information to developers in three major sections: First, preliminary information including the report title, device, and short description.  Second, a list of the steps with the following information regarding each step is displayed: (i) the action for each step, (ii) the type of a component, (iii) the relative location of the component, (iv) the \texttt{activity} Java class where the component is instantiated in the source code, and (v) the component specific screenshot.  Third, a list of full screen-shots corresponding to each step is presented at the bottom of the page so the developer can trace the steps through each application screen. 

\vspace{-0.20cm}
\subsection{Preliminary Results}

	To evaluate \Fusionsp (See full details in \cite{Moran:FSE2015}) we investigated its ease of use, as well as the reproducibility of the \Fusionsp reports compared to reports created using Google Code Issue Tracker (GCIT).  First, in the \textit{bug-creation study} we recruited eight students (four undergraduate or \textit{non-experts} and four graduate or \textit{experts}) to construct bug reports using \Fusionsp and GCIT --- as a representative of traditional bug tracking systems--- for 15 real-world world bugs in 14 open-source apps from F-Droid.  We collected survey responses from these participants regarding the ease of use and user preferences of each tool.  Next, in the \textit{bug-reproduction study} we evaluated the reproducibility of the \Fusionsp and GCIT reports generated by the first group of participants. These reports (120 for each type) and the original bug reports extracted from the respective app issue trackers were evaluated by a new set of 20 graduate student participants through attempted bug reproduction on physical devices. The results for the \textit{bug creation study} indicate that \textit{\Fusionsp was about as easy to use as the GCIT for experienced participants but was more difficult for inexperienced participants to use compared to GCIT.} The results for the \textit{bug-reproduction study} clearly illustrate that \textit{developers using FUSION are able to reproduce more bugs compared to traditional bug tracking systems such as the GCIT.}
\vspace{-0.30cm}
\section{Proposed Research Program}
\label{sec:research-program}
 The information gap and challenges outlined in Section \ref{sec:intro} point to three major challenegs that a research program must address to transform the state of bug reporting: (i) provide bug reports to developers with immediately actionable knowledge (reliable reproduction steps, stack traces and replayable scenarios), (ii) facilitate reporters providing this information through the design and implementation of new reporting mechanisms, and (iii) illustrate the value of the reports created in improving the performance of state-of-the-art techniques to support the costly tasks of developer triaging and duplicate bug report detection. We propose the following research tasks to address these challenges. 
 	
	\textit{(1)Design and implement a comprehensive framework of bug reporting mechanisms that enables on-device, off-device, and automatic creation of structured, detailed issue reports with low user effort.} The first of these mechanisms will constitute an off-device, web-based reporting system that facilitates users constructing detailed bug reports with reproduction steps using autocompletion, driven by program analysis and a statistical \textit{n}-gram model of an app's event-flow (leveraging experience from prior work \cite{Linares:MSR15}).  The second mechanism is an on-device reporting mechanism that will use a smart-capture and replay system to allow users to record bugs in-app, directly on their devices, while preserving user privacy.  In addition to these manual reporting approaches, there is a growing need for automated fault reporting as the development of automated oracles for mobile applications advances, and no current automated testing solutions support the detailed reporting of uncovered faults.  The third mechanism is an automated reporting mechanism for app crashes.  This automated approach systematically exercises an app's GUI according to several strategies informed by static analysis, with the intrinsic goal of uncovering crashes and generating detailed crash reports\cite{Moran:ICST16}. \textit{(2) Enhance the quality and lower the cost of developer triaging and duplicate bug report detection  by designing new approaches to support these tasks that fully leverage the information contained within reports created with our framework.}  We plan to develop new techniques that leverage the structured information in our novel reports, in order to assign reports to relevant developers who most easily be able to resolve them, and eliminate the large numbers of duplicate reports. Additionally, we envision experimenting with techniques that utilize our automated program analysis techniques in order to evolve bug reports and associated test cases as an app evolves. {\textit{(3) Design studies to empirically evaluate (i) the user experience of reporting mechanisms, (ii) the improvements in readability and reproducibility of resulting reports and (iii) the performance enhancements of existing techniques for supporting duplicate bug report detection and developer triaging.}  We plan to carry out extensive evaluations of the reporting mechanisms and applications of the reports to SE tasks using popular open source apps from F-Droid and Google Play.  These evaluations will take the form of user studies to evaluate the reporting mechanisms and empirical case studies to evaluate the applications.

\vspace{-0.30cm}
\section{Anticipated Contributions}
\label{sec:contributions}
This research proposal addresses the resource-consuming and often costly tasks of bug report creation, reproduction, and fixing for mobile and GUI-based apps representing a paradigm shift in the manner that developers and SE researchers perceive the process of app issue reporting.  In addition to the contributions in terms of tools and novel applications of program analysis and statistical modeling,  we anticipate the potential impact of this research program to be immediate and far-reaching in terms of providing actionable tools for developers, and spurring new research directions related to defect reporting and fixing in the SE community.  For instance, today developers rely on the poorly structured system of textual user reviews and simple ratings to engage their users, estimate the success of their app, and uncover potential defects.  The reporting techniques in this proposal could potentially be integrated into such reviewing systems, giving users a familiar, intuitive option to give feedback to developers regarding defects and feature requests. Furthermore, the proposed approach addresses challenges specific to mobile bugs. Most notably, by combining traditional natural language bug reports with high-level adaptable execution scenarios, the reports carry the potential to be automatically replayed and confirmed on several different hardware and device configurations and updated as an app evolves.   \textit{We predict that the results of this research program, if successful, will force researchers and practitioners to rethink how bug reports are created and used in the software development and maintenance process}.
\vspace{-0.25cm}

\section{Conclusion}
\label{sec:concl}
We present a comprehensive research agenda, bolstered by strong preliminary work, focused on improving the state of the art and practice in reporting bugs for mobile and GUI-based applications. This work has the potential to challenge current thinking regarding how bug reports are constructed and applied toward software maintenance tasks.  
\vspace{-0.3cm}

\balance
\small{
\bibliography{ms}
\bibliographystyle{abbrv}}
\balancecolumns

\end{document}